\documentclass[12pt]{article}
\usepackage{graphicx}

\begin{document}
\title{Improved Methods for Making Inferences About Multiple Skipped Correlations}
\author{Rand R. Wilcox \\
Dept of Psychology \\
University 
of Southern California\\
\\
Guillaume A. Rousselet\\
Institute of Neuroscience and Psychology,\\
College of Medical, Veterinary and Life\\
University of Glasgow\\
\\
Cyril R. Pernet\\
Centre for Clinical Brain Sciences, Neuroimaging Sciences\\
University of Edinburgh\\
}
\maketitle

 

\begin{center}
ABSTRACT
\end{center}
A skipped correlation has the advantage of dealing with outliers in a manner that takes into account
the overall structure of the data cloud. For p-variate data, $p \ge 2$,
there is an extant method for testing the hypothesis of a zero correlation for each pair of variables that is
designed to control 
 the probability of one or more
Type I errors. And there are methods for the related situation where the focus is on
the association between a dependent variable and $p$ explanatory variables. However,
there are limitations and several concerns with extant techniques. 
The paper describes alternative approaches that deal with
these issues.


\section{Introduction}
For  some unknown 
$p$-variate distribution, let $\tau_{jk}$
be some measure of association between variables $j$ and 
$k$, $1\le j < k \le p$. 
A basic goal is controlling the family wise error rate (FWE), 
meaning the probability of at least one Type I error, when
 testing
\begin{equation}
H_0: \tau_{jk}=0   \label{cor.test}    
\end{equation}
for each $j<k$, where the alternative hypothesis is $H_1$: $ \tau_{jk} \ne 0$.

Of course, one possibility is to take $\tau_{jk}$ to be Pearson's correlation.
It is well known, however, that Pearson's correlation is not robust (e.g., Wilcox, 2017a). 
In particular, it has an unbounded influence function (Devlin et al., 1981). Roughly, 
this means that if one of the marginal distributions is altered slightly,
 the magnitude of $\rho$  can be changed substantially.
A related concern is that the usual estimate
of $\rho$, $r$, has a breakdown point of only $1/n$.  
This means, for example, that even a single outlier, properly placed, can 
completely dominate  $r$. 

Let $X_1, \ldots, X_p$ denote $p$ random variables.
As is evident, a goal related to methods design to test  (1) is to test
\begin{equation}
H_0: X_j  \, {\rm and} \, X_k \, {\rm are\,  independent}   \label{ind.test}
\end{equation}
for each $j<k$, where the alternative hypothesis is that there is some type of dependence.
 Note that when a method is based on an estimate of $\tau_{jk}$, if it assumes 
homoscedasticity (the variance of $X_j$ does not depend on the value of $X_k$), 
 it can detect dependence that would be missed by a method designed
 to test (1) that is
insensitive to heteroscedasticity. Consider, for example, the classic Student's t-test for testing (1) based on 
Pearson's correlation,  $\rho_{jk}$. Even when $\rho_{jk}=0$, if there is heteroscedasticity, the probability 
of rejecting increases as the sample size gets large (e.g., Wilcox, 2017b). The reason is that the test statistic, T,
 uses the wrong standard error. So if the goal is to detect dependence, T might be more effective than
a method that is insensitive to heteroscedasticity. But if the goal is to compute a confidence interval
for  $\tau_{jk}$, methods that assume homoscedasticity can be highly unsatisfactory.
The goal here is to consider both situations: testing (2) via some method based on an estimate of $\tau_{jk}$ as well
as a method for testing (1) that allows heteroscedasticity. 

Consider the usual linear regression model,  where $X_1, \ldots, X_p$ are $p$ predictors and $Y$ is some
dependent variable of interest.
Let $\tau_{yj}$ be some measure of association associated with $Y$ and $X_j$
($j=1, \ldots, p$.)   There is, of course, a goal related to (1), namely testing
\begin{equation}
H_0: \tau_{yj}=0    \label{reg.assoc}
\end{equation}
for each $j$ in a manner that controls FWE. Relevant results are included here.

Note that both (1) and (2) are not based on a linear model in the sense that  no single variable is designated as the
dependent variable with the remaining $p-1$ variables viewed as the explanatory variables. 
In the context of a linear model, an alternative to (3), with the goal of detecting dependence, is to test
\begin{equation}
H_0: \beta_j=0 
\end{equation}
for each $j$  ($j=1, \ldots, p$),
  using some robust regression estimator,
where $\beta_j=0$ is the slope associated with the $j$th explanatory variable.
From basic principles, the magnitude of the slopes can depend on which explanatory variables are included in the model.
In particular, the magnitude of any slope can depend on the nature of the association between the corresponding 
explanatory variable and the other explanatory variables in the model. 
In contrast, the method used here to test (1) is not impacted in the same manner as will be made evident.
So 
in terms of power, testing (1) can have more or less power than testing (4).
  
 Now consider the situation where say $H_0$:  $\beta_1=0$ is tested when the remaining explanatory variables are
 ignored. One could use some robust regression  estimator to accomplish this goal.
  However, robust regression estimators can react differently to outliers, compared to the robust regression correlation
  used here. Details are described and illustrated in section 4.4 of this paper.

Numerous  robust
 measures of association have been proposed 
that belong to one of two types (e.g., Wilcox, 2017a).
The first, sometimes labeled Type M, are measures that guard against the deleterious impact of outliers among
the marginal distributions. The two best-known Type M measures are Spearman's rho and Kendall's tau. A positive feature
is that both have  a bounded influence function (Croux \& Dehon, 2010).  Nevertheless, two outliers, properly placed,
can have in inordinate impact on the estimate of these measures of association. 
In fact, imagine that boxplots are used to detect  outliers among the first of two 
random variables, ignoring the second random variable, and none are found. Further imagine that
no outliers among the second random variable are found, ignoring the first random variable.
As illustrated in Wilcox (2017b, p. 239), it is still possible that there are outliers relative to the cloud of points that
can have an inordinate impact on Spearman's rho and Kendall's tau.
More generally, methods that deal only with outliers among each of the marginal distributions can be unsatisfactory.

Type O measures of association are designed to deal with the concern with 
Type M measures of association just described. A simple way to proceed 
 is to use a skipped measure
of association. That is, 
 use a multivariate outlier detection method that takes into account
the overall structure of the data cloud,   remove any points flagged
as outliers, and compute a measure of association (e.g., Pearson's correlation) 
based on the remaining data. Skipped correlations are special cases of a general approach
toward multivariate measures of location and scatter suggested
by Stahel (1981) and Donoho (1982).

There are several outlier detection techniques that take into account the overall structure of a data cloud
(e.g., Wilcox, 2017a, section 6.4). 
Consider a random sample of $n$ vectors  from some multivariate distribution: $\mathbf{X}_{i}$
($i=1, \ldots, n$).
One general approach is to measure the distance of the point  $\mathbf{X}_{i}$ from the center of the cloud of data  with 
\[D_i = \sqrt{({\bf X}_i - {\bf C})^{\prime}
{\bf M}^{-1}({\bf X}_i - {\bf C})},\]
where $\mathbf{C}$ and $\mathbf{M}$ are robust measures of location and scatter, respectively, having reasonably high breakdown points.
Among the many 
possible choices  for  $\mathbf{C}$ and $\mathbf{M}$
are the minimum volume ellipsoid estimator (Rousseeuw \& van Zomeren, 1990), the minimum covariance 
determinant estimator (Rousseeuw \& van Driessen,1999), the TBS estimator proposed by Rocke (1996).
If $D_i$ is sufficiently large, $\mathbf{X}_{i}$ is flagged an outlier.  There are several refinements of this approach, but the details
go beyond the scope of this paper.

Here, following Wilcox (2003), a projection-type method is used, 
 which is  reviewed in section 2. 
 The basic 
 idea is that if a point is an outlier, then it should be
an outlier for some
projection of the $n$  points.
This is not to suggest that it dominates all other methods. 
The only suggestion is that it is a reasonable choice with the understanding that there might be
practical advantages to using some other technique. It will be evident that the approach used here
 is readily generalized to any measure of association that might be used in conjunction with any 
multivariate outlier detection technique that might be of interest. 

Now, consider the strategy of removing points flagged as outliers using a projection method and then computing
some measure of association based on the remaining data, say
 $\hat{\tau}_{jk}$.
 Wilcox (2003) describes a method for testing (1) when using 
 Spearman's rho.
However, there are limitations and concerns that motivated this paper. First,
a simple method for determining a 
0.05 critical value for the test statistic, used by Wilcox,  is available. In principle, simulations could be used
to determine a critical value when testing at say the 0.01 level, but the efficacy of doing this is unknown.
 Second, 
when sampling form heavy-tailed distributions, the method avoids FWE greater than 0.055  
in simulations reported by Wilcox (2003), but under normality 
the actual probability of one or more Type I errors can drop substantially below the nominal
level. What would be desirable is a method that controls FWE in a manner that is less sensitive to changes
in the distribution generating the data. Another issue that might be a concern is that
the  method is limited to using Spearman's rho. It breaks down when using Pearson's correlation instead.
The goal here is to describe alternative approaches that deal with all of these limitations.

For $p=2$, Wilcox (2015) found that a percentile bootstrap method performed well
in simulations,  in terms of avoiding a Type I 
error probability greater than 0.05, when testing (1) at the 0.05 level, even when there is heteroscedasticity.  So 
when $p>2$ a simple approach is to use this method in conjunction with
 some extant technique for controlling the probability of one or more Type I errors.
However, as will be demonstrated, the actual level can be substantially smaller than the
nominal level when the sample size is relatively small, which indicates that power can be 
relatively low. A method for dealing with this concern is derived.

The paper is organized as follows. Sections 2 and 3 review
the approach used by Wilcox (2003). Section 2 describes  the details of the outlier technique
that was used
and 
Section 3 reviews the method used
to test (1). Section 4 describes  the proposed methods.
Section 5 reports simulation results and  section 6 illustrates the methods using data
from two studies. The first deals with the reading ability of children and the second
deals with the processing speed in adults.

\section{A Projection-Type Outlier 
Detection Method}

The multivariate outlier detection technique used by Wilcox (2003) is computed as follows.
Let 
$\hat{\xi}$ be some robust multivariate measure of location. Here, $\hat{\xi}$ is taken to be the marginal medians,
but there is a collection of alternative estimators that might be used (e.g., Wilcox, 2017a, section 6.3).
Let $\mathbf{X}_1, \ldots, \mathbf{X}_n$ be a random sample from some multivariate distribution.
Fix $i$ and for the point $\mathbf{X}_i$, project all $n$ points onto the line
connecting $\hat{\xi}$ and $\mathbf{X}_i$ and let $D_{j}$ be the
distance between the origin and  the projection of  $\mathbf{X}_j$.
More formally, let 
\[A_i = \mathbf{X}_i - \hat{\xi},\]
\[B_{j} = \mathbf{X}_{j} - \hat{\xi},\]
where both $A_i$ and $B_j$ are  column vectors having length $p$,
and let
\[C_j = \frac{A^{\prime}_iB_j}{B^{\prime}_jB_j}B_j,\]
$j=1, \ldots, n$.
Then when projecting the points onto the 
line between $X_i$ and $\hat{\xi}$, the distance of the
$j$th projected point from the origin is
\[D_j = \|C_j\|,\]
where $\|C_j\|$ is the Euclidean norm of the vector $C_j$.

Next, a modification of the boxplot rule for detecting outliers
is applied to the $D_j$ values,  which has close similarities
to one used by Carling (2000).
Let $\ell = [n/4 + 5/12]$, where $[.]$
is the greatest integer function,
let 
\[h=\frac{n}{4} +\frac{5}{12}- \ell\]
and $k=n-\ell+1$.
Let $D_{(1)} \le \cdots \le D_{(n)}$ be the $n$ 
distances written in ascending order.
The so-called ideal fourths associated with the
$D_j$ values are
\[q_1 = (1-h)D_{(\ell)} + hD_{(\ell+1)}\]
and
\[q_2 = (1-h)X_{(k)} + hX_{(k-1)}.\]
Then the jth point is declared an outlier if
\begin{equation}
 D_j > M_D + \sqrt{\chi^2_{0.95,p}}(q_2-q_1),
\end{equation}
where $M_D$ is the usual sample median based on the $D_j$ values
and $\chi^2_{0.95,p}$ is the 0.95 quantile of a chi-squared
distribution with $p$ degrees of freedom (cf. Rousseeuw and van Zomeren, 1999).

The process just described is for a single
projection; for fixed $i$, points are projected onto the line
connecting $\mathbf{X}_i$ to $\hat{\xi}$.
Repeating this process for each
$i$, $i=1, \ldots,  n$, a point is
declared an outlier if for any of these projections, 
it satisfies equation (4).

A simple and seemingly desirable
modification of the method just described 
is to replace the interquartile range
($q_2-q_1$) with the median absolute deviation (MAD) measure of scale based
on the $D_j$ values. So here, MAD is the median of the values 
\[|D_1 - M_D|, \ldots, |D_n - M_D|.\]
Then the jth point is declared an outlier if
\begin{equation}
 D_j > M_D + \sqrt{\chi^2_{0.95,p}} \frac{{\rm MAD}}{0.6745},
\end{equation}
where the constant 0.6745 is typically used because under
normality, MAD/0.6745 estimates the standard deviation.
One appealing feature of MAD is that it has a higher
finite sample breakdown point versus the interquartile range.
MAD has a finite sample breakdown point of approximately 0.5, while
for the interquartile range it is only 0.25. 
Let $p_n$ be the outside rate per observation
corresponding to some outlier detection method, which is the
expected proportion of outliers based on a random sample of size
$n$.  
A negative feature associated with  (5)
is that $p_n$ appears to be considerably less stable
as a function of $n$. In the bivariate case,
for example, it is approximately 0.09 with $n=10$ and drops
below 0.02 as $n$ increases. 
For the same situations, $p_n$ based on equation 
(4) ranges between 0.043 and 0.038. Perhaps situations are encountered where the higher
breakdown point associated with (5) is more important than having $p_n$ relatively stable as a function
of the sample size $n$. But for present purposes, 
 the approached based on (5)
is not used.

\section{A Review of an Extant Technique}

Momentarily  consider the bivariate case and let $r_c$ be skipped correlation when
 Pearson's correlation is used after
outliers are removed as described in section 2. 
An unsatisfactory approach to testing (1) is to simply use the test statistic 
\[T=r_c \sqrt{\frac{m-2}{1-r^2_c}},\]
 where $m$ is the number of
observations left after outliers are discarded, and reject
if $|T|$ exceeds the $1-\alpha/2$ quantile of Student's T distribution
with $m-2$ degrees of freedom. 
This approach is unsatisfactory, even under normality, roughly because
the wrong standard error is being used. Wilcox (2003) illustrated that
if this issue is ignored and this approach is used anyway, it 
results in  poor control over the probability of a Type I error.

Returning to the general case $p \ge 2$,
Wilcox (2003) proceeded as follows given the goal of controlling
 FWE.
Let $\hat{\tau}_{cjk}$
be Spearman's correlation between variables $j$ and $k$
after outliers are removed. 
Let
\[T_{jk} = \hat{\tau}_{cjk}  \sqrt{\frac{n-2}{1-\hat{\tau}_{cjk}^2}},\]
and let 
\begin{equation}
T_{{\rm max}} = {\rm max} |T_{jk}|,
\end{equation}
where the maximum is taken overall $j<k$.
The initial strategy was to estimate, via simulations,
the distribution of $T_{{\rm max}}$
under normality when all correlations are zero and $p < 4$, determine
the 0.95 quantile, say $q$, for 
$n=10$, 20, 30, 40, 60, 100 and 200,
and then reject $H_0: \tau_{cjk} =0$
if $|T_{jk}| \ge q$. 
However, for 
 $p \ge 4$, this approach was unsatisfactory when dealing with
 symmetric, heavy-tailed distributions, roughly meaning that outliers are relatively common.
 More precisely, the estimate of FWE exceeded 0.075.
So instead, the critical value $q$ was determined via simulations where data are generated from 
 a g-and-h distribution (described in the next section) with
$(g,\, h) = (0, \, 0.5)$, which is a symmetric and heavy-tailed distribution. This will be called method M henceforth. 
Using instead Pearson's correlation, the method just described did not perform well in simulations.
Moreover, 
there are at least three practical concerns with method  M, which were reviewed in the
introduction.

\section{The Proposed Methods}

This section describes a method for testing (2), based on a skipped correlation, which is
sensitive to heteroscedasticity as well as  the extent to which $\tau_{jk}$ differs from zero.
This is followed by  methods for  testing (1) that is designed to control
FWE even when there is heteroscedasticity. Results dealing with (3) are described as well. 

\subsection{A Homoscedastic Method for Testing (2)}

An outline of the proposed method for testing (2) is as follows.
Let $X_{ij}$ ($i=1, \dots, n$; $j=1, \ldots, p$) be a random sample of $n$ vectors from a $p$-variate distribution.
The basic idea is to generate bootstrap samples from each marginal distribution in a manner for which 
there is no association. Next, compute $T_{{\rm max}}$ based on these bootstrap samples yielding say 
$T^*$. This process is repeated $B$ times, which can be used to estimate the $1-\alpha$ 	quantile of the distribution
of  $T_{{\rm max}}$ when the null hypothesis is true for all $j<k$. 

To be more precise, let $X^*_{11}, \ldots, X^*_{n1}$ be a bootstrap sample from the first column of the data matrix, which
is obtained by randomly sampling with replacement $n$ values from $X_{11}, \ldots, X_{n1}$. Next,
take an independent bootstrap sample from the second column yielding $X^*_{12}, \ldots, X^*_{n2}$ and continue in this
manner for all $p$ columns. So information about the marginal distributions is maintained but the bootstrap
samples are generated in a manner so that all correlations are zero.
Next, compute $T^*$, the value of $T_{{\rm max}}$ based on these $p$ bootstrap samples and repeat this
process $B$ times yielding $T^*_1, \ldots, T^*_B$. Here, $B=500$ is used, which generally seems to suffice,
in terms of controlling the  probability of a Type I error, when
dealing with related situations (Wilcox, 2017a).

Next, put the $T^*$ values in ascending order yielding $T^*_{(1)} \le \cdots \le  T^*_{(B)}$.
Let $c=(1-\alpha)B$, rounded to the nearest integer. Then an estimate of the $1-\alpha$ quantile of the null
distribution of $T_{{\rm max}}$  is $T^*_{(c)}$. However, simulations indicated that the
Harrell and Davis (1982) quantile estimator performs  a bit better and so it is used henceforth.  
That is, the $q$th quantile is estimated
with 
\[ \hat{\theta}_q =  \sum W_b T^*_{(b)},\]
where 
\[W_b = P\left(\frac{b-1}{B} \le U \le \frac{b}{B}\right),\]
and $U$ has a beta distribution with parameters $(B+1)q$ and $(B+1)(1-q)$. 
So if the goal is to have the probability of one or more Type I errors equal to 
$\alpha$, then for any $j<k$, reject (2) if $|T_{jk}| \ge  \hat{\theta}_{1-\alpha}$. 

Notice that an analog of a p-value can be computed. That is, one can determine the smallest $\alpha$ value for
which one or more of the $(p^2-p)/2$ hypotheses is rejected. Here this is done by finding the value $q$ 
that minimizes $| \hat{\theta}_q  - T_{{\rm max}}|$.
When using Spearman's rho after outliers are removed, this will be called method SS henceforth. Using 
Pearson's correlation will be called method SP. 

Note that methods SS and SP  are based on a nonparametric
estimate of the marginal distributions. However,  the estimate of the $1-\alpha$ quantile of the null
distribution of $T_{{\rm max}}$  
is done for a situation where there is homoscedasticity. 
That is, based on how the bootstrap samples were generated, any information regarding how the
variance of $X_j$ depends on the value of $X_k$ is lost. 
If there is in fact heteroscedasticity, meaning that the
variance of $X_j$ depends on the value of $X_k$,
and if the goal is to test (1), not (2), methods SS and SP can be unsatisfactory based on simulations reported
in section 5. For example, when testing at the 0.05 level, the actual FWE can exceed 0.10.

\subsection{A Heteroscedastic Method for Testing (1)}

Results in Wilcox (2015) hint at how one might proceed when testing (1). Focusing on $p=2$, he found
that when a skipped correlation based on Pearson's correlation, a basic percentile bootstrap method performs reasonably well in terms of 
avoiding a Type I error probability greater than  0.05 when testing at the 0.05 level.
So in contrast to methods SS and SP, a bootstrap sample is obtained by sampling with replacement
$n$ rows from the $n$-by-$2$ matrix $X_{ij}$. Next, compute  the skipped correlation based on this
bootstrap sample and label the result $\hat{\tau}^*$. Repeat this process $B$ times yielding
$\hat{\tau}^*_{1}, \ldots, \hat{\tau}^*_{B}$.  
Put these $B$ values in ascending order yielding $\hat{\tau}^*_{(1)}, \ldots, \hat{\tau}^*_{(B)}$.
Let $\ell= \alpha B/2$,  rounded to the
nearest integer and $u=B- \ell$. Then the $1-\alpha$ confidence interval for $\tau_{jk}$ is taken to be
\begin{equation}
(\hat{\tau}^*_{(\ell+1)}, \, \hat{\tau}^*_{(u)}).
\end{equation}
Letting $Q$ denote the proportion of $\hat{\tau}^*_{b}$ ($b=1, \ldots, B$) values less than zero, a (generalized)
p-value is given by 2min\{$Q$, 1-$Q$\}. (For general theoretical results related to this method, see Liu \& Singh,
1997.)
 The striking feature of the method is that
there is little variation in the estimated Type I error probability among the situations considered in the simulations.
For $n=40$, the estimates ranged between 0.021 and 0.030. For $n=20$, estimates are less than 
0.020.

The results just summarized
 suggest a simple modification for dealing with $p \ge 2$ and for $\alpha \ne 0.05$: 
 use simulations to estimate a critical p-value, $p_{\alpha}$,  such that FWE is equal to $\alpha$ under
normality and homoscedasticity.  
To elaborate, for any $j<k$,  let $p_{jk}$ be the percentile bootstrap 
p-value when testing (1). 
Then proceed as follows:
\begin{itemize}
\item Generate $n$ observations from a p-variate normal distribution for which the covariance matrix is equal to the
identity matrix.

\item For the data generated in step 1, compute the p-value for each of the $C=(p^2-p)/2 $ hypotheses using 
a percentile bootstrap method. Let  $V$ denote the minimum p-value among the $C$ p-values just
computed. 
\item Repeat steps 1 and 2 $D$ times yielding the minimum p-values $V_1, \ldots, V_{D}$.

\item Let $\hat{p}_{\alpha}$ be an estimate of $\alpha$  quantile of the distribution of  $V$. Here the 
Harrell and Davis  estimator is used.

\item Reject any hypothesis for which $p_{jk} \le \hat{p}_{\alpha}$.

\end{itemize}

Here, 
$D=1000$, which
was motivated in part by  the execution time  required to estimate  $p_{\alpha}$.
Using a four quad Mac Book Pro with 2.5 GHz processor, execution time is about 15 minutes when
$n=20$ and $p=4$. (An R function was used that takes advantage of multicore processor via the R package
parallel.) Increasing the sample size to $n=40$, execution time exceeds 36 minutes.
This will be called method ECP henceforth.

\vspace*{5mm}
 {\bf Method H}
  \vspace*{5mm}

Note that FWE could be controlled using the method derived by Hochberg (1988), which is applied as follows.
Compute a p-value for each of the $C$ tests to be
performed and label them $P_1, \dots, P_C$. Next, put the p-values
 in descending order yielding
$P_{[1]} \ge P_{[2]} \ge \cdots \ge P_{[C]}$.
Proceed as follows:
\begin{itemize}

\item Set k = 1.
\item
If $P_{[k]} \le d_k$, where $d_k=\alpha/k$,
reject all $C$ hypotheses; otherwise, go to step 3.

\item
Increment $k$ by 1.
If  $P_{[k]} \le  d_k$, stop and reject all hypotheses having
a p-value less than or equal to $d_k$.

\item If $P_{[k]} >  d_k$, repeat step 3.

\end{itemize}
This will be called method H henceforth.

Method H avoids the high execution time  associated with estimating the 
critical p-value,  $p_{\alpha}$, used by method ECP. But with a relatively
small sample size, this approach will have
 relatively low power roughly because  the actual FWE can be considerably less than the nominal level.
 This was expected because as previously noted, when $p=2$, the actual level can be
 considerably less than the nominal level.
 For example, with $p=4$ and $n=20$, FWE is approximately 0.007 under normality and homoscedasticity when
 the nominal level is 0.05. A modification of method H that deals with this issue, which also avoids the excessively high execution time due to estimating $p_{\alpha}$, is
 described in section 4.3.
  There are other methods for controlling FWE that are closely related to Hochberg's method
 (e.g., Wilcox, 2017, section 12.1), but it is evident that 
 again the actual level can be substantially smaller than the nominal level.   
 
 \vspace*{5mm}
 {\bf Method L}
  \vspace*{5mm}
  
 Observe that both methods H and ECP are readily extended to testing (3). 
 Now a skipped correlation refers to the strategy of removing outliers 
   based on $(Y_i, \mathbf{X}_{i})$  ($i=1, \ldots, n$)
   and computing some measure of association using the
   remaining data. 
   The modification of ECP, aimed at testing (3), is called method L henceforth.
    Another approach is, for each $j$, remove outliers among
   $(Y_i, X_{ij})$, ignoring the other $p-1$ independent variables, and then 
   compute a measure of association ($j=1, \ldots, p$). 
   Perhaps there is some practical advantage to this latter
   approach, but it substantially increases execution time, so it is not pursued.  
    
 \subsection{Improving Methods H and L: Methods L3 and H1}
 
 This section describes a 
 modification of method H aimed at testing (1) and method L aimed at testing (3)
 that avoid 
the high  execution time associated with method ECP. 
First consider testing (\ref{reg.assoc}) via method L
 in conjunction with Pearson's correlation. 
 Momentarily focus on the case of a single
 independent variable ($p=1$). Simulations indicate that $p_{\alpha}$ slowly decreases to $\alpha$ as $n$ increases.
 For $\alpha=0.05$ and $n=30$, $p_{\alpha}$ is estimated to  0.087. For $n=80$, 100 and 120 the
 estimates are  0.076, 0.062 and 0.049, respectively.
So  the basic idea is to use simulations to
 estimate the distribution of  the  p-value for a few selected sample sizes 
and use the results  to get an adjusted p-value. 
The estimate is based on the approach used  in conjunction with method  
 ECP, only now $D=2000$ replications are used.  Preliminary simulations suggest using
 estimates for $n=30$, 60, 80 and 100. 

To elaborate (still focusing on $p=1$) 
let $V_n$ denote the vector 
 of  p-values  corresponding to $n$ and stemming from the simulation just described. 
 Let $p_{y1}$ be the bootstrap p-value when testing  (\ref{reg.assoc}) via method L.
 Then given $V_n$, an adjusted p-value can be computed, which 
  is simply the value $q$ such that $\hat{\theta}_q(V_n)=p_{y1}$. 
 For example, if the bootstrap p-value is 0.08, and $V_n$ indicates that the level of the
 test is 0.05 when $p_{\alpha}=0.08$ is used, then the adjusted p-value is 0.05.
Here, adjusted p-values are computed in the following manner:
 use $V_{30}$ when $20 \le  n \le 40$, use $V_{60}$ when $41 <n \le  70$,
 use $V_{80}$ when $71 < n  \le 100$ and $V_{100}$ when $101 < n  \le 120$. For $n>120$, 
 no adjustment is made. 
 
 Now consider  $p>1$ and for each $j$ ($j=1, \ldots, p)$, let $\tilde{p}_j$ be the adjusted p-value
when  testing (\ref{reg.assoc}) via method L.
 The strategy is to  use these
  adjusted p-values in conjunction with Hochberg's method to control FWE.
 This will be called method L3.
 As is evident, the same strategy can be used when testing (1); this will be
   called method H1 henceforth.
   
   \subsection{Comments on Using a Robust Regression  Estimator}

Now consider the usual linear model involving a single explanatory variable and focus on the goal  of testing $H_0$:  $\beta_1=0$.
There are several robust regression estimators that might be used that have a high breakdown point.
   They include the M-estimator derived by Coakley and Hettmansperger (1993),  the MM-estimator derived by Yohai   (1987), 
as well as the Theil (1950) and Sen (1964) estimator. 
The first two estimators have the highest possible breakdown point, 0.5. The breakdown point  of the Theil--Sen estimator
is approximately 0.29. 
Currently, a percentile bootstrap method appears to be a relatively good technique for testing  $H_0$:  $\beta_1=0$ when
dealing with both non-normality and a heteroscedastic error term (Wilcox, 2017a). 
 But despite the relatively high breakdown points, a few outliers
among $(Y_1, X_1), \ldots, (Y_n, X_n)$ can have an inordinate impact on the estimate of the slope as illustrated in Wilcox (2017a, section 10.14.1).
Moreover, the impact of outliers on any of these robust regression estimator can differ substantially from the
impact on the skipped correlation, which can translate into differences in power.

As an illustration, consider the situation where
$Y=X+ \epsilon$, where $\epsilon$ has a standard normal distribution,
the sample size is 
 $n=30$ and  two outliers are introduced by setting $(X_1, Y_1)=(X_2, Y_2)=(2,1, -3.4)$.  Power using
 methods L is 0.53  (based on a simulation with 2000 replications) compared to 0.33 using the
MM-estimator. The reason is that the skipped correlation is better able to detect and eliminate the two outliers. 
But this is not to suggest that method L dominates in terms of power. When using the MM-estimator,
outliers can inflate the estimate of the slope  resulting in more power. 
The same issue arises using the other robust regression estimators previously listed---evidently no method
dominates in terms of power. The only certainty is that the choice of method can make a practical difference.  


\section{Simulation Results}

Simulations were used to check the small sample  properties of 
method  SS and SP for the exact same situations used by Wilcox (2003).
Observations were generated where the marginal distributions
 are independent with each having 
 one of four  g-and-h distributions
(Hoaglin, 1985),
which contains the standard  normal distribution as a special case.
If $Z$ has a standard normal distribution, then
\begin{equation}
W = \left\{ \begin{array}{ll}
 \frac{{\rm exp}(gZ)-1}{g} {\rm exp}(hZ^2/2), & \mbox{if $g>0$}\\
  Z{\rm exp}(hZ^2/2), & \mbox{if $g=0.$}
   \end{array} \right. 
\end{equation}
has a g-and-h distribution where $g$ and $h$ are parameters that 
determine the first four moments.
The four distributions used here were the standard normal ($g=h=0$), a
symmetric heavy-tailed distribution ($h=0.5$, $g=0$), an asymmetric 
distribution with 
relatively light tails ($h=0$, $g=0.5$), and an asymmetric distribution with 
heavy tails ($g=h=0.5$). 

Table 1 shows the theoretical skewness and kurtosis
for each distribution considered. When $g>0$ and 
$h>1/k$, $E(W^k)$ is not defined and
the  corresponding entry in Table 1 is left blank.
Plots of the four distributions are shown in Figure 1.
Additional properties of the g-and-h distribution
are summarized by Hoaglin (1985).

\begin{table}
\caption{Some properties of the g-and-h distribution.}
\centering
\begin{tabular}{ccrr} \hline
g & h &  $\kappa_1$ & $\kappa_2$\\ \hline
0.0 & 0.0  &  0.00 & 3.0\\
0.0 & 0.5 & --- & ---\\
0.5 & 0.0  & 1.75 & 8.9\\
0.5 & 0.5 &  --- &  ---\\ \hline
\end{tabular}
\end{table}

\begin{figure} 
\resizebox{\textwidth}{!}
{\includegraphics*[angle=0]{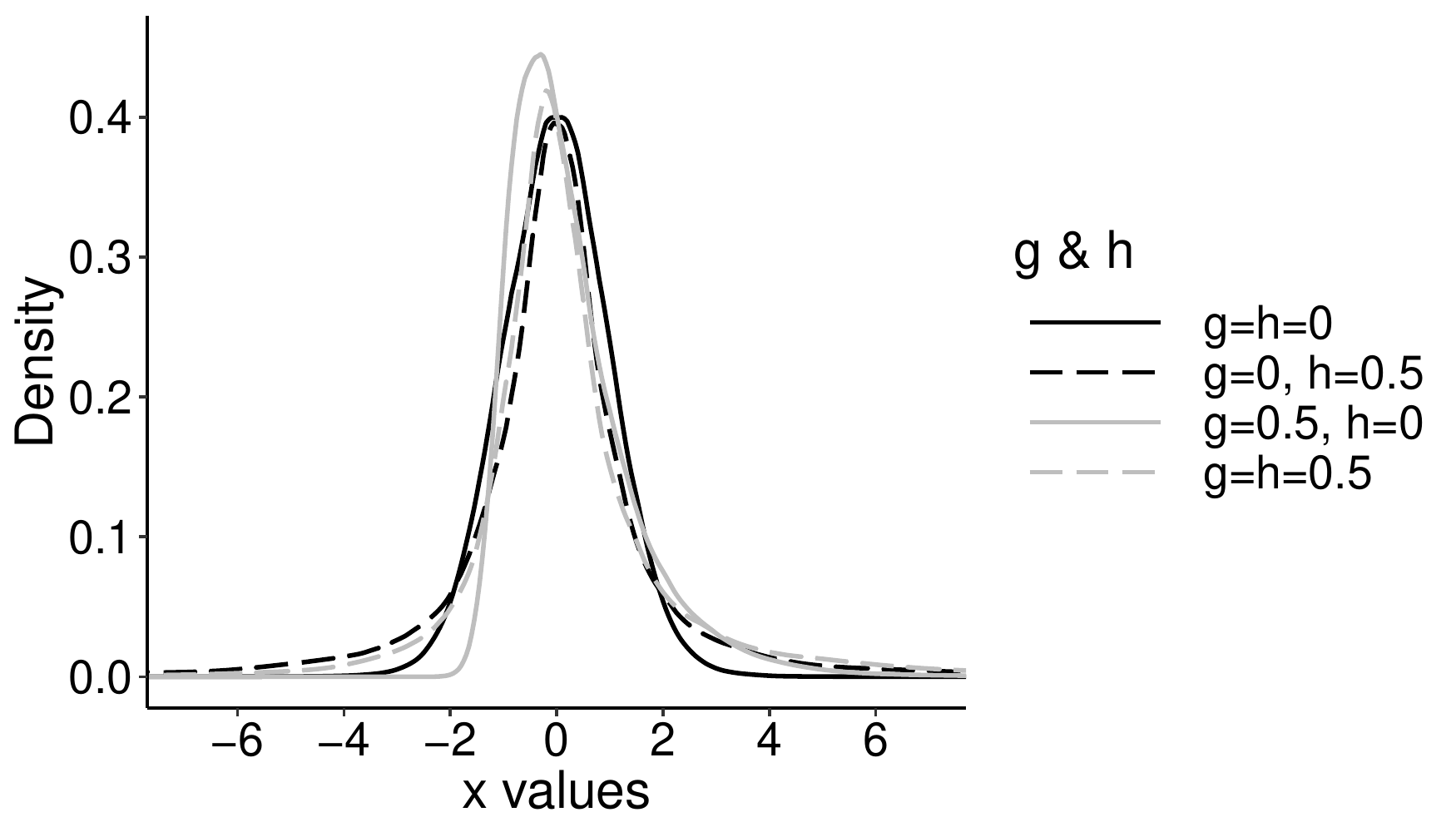}}
\caption{The four g-and-h distributions used in the simulations.}
\end{figure}


First consider methods SS and SP for testing (2).
Each replication in the simulations consisted of generating 
$n$ observations and 
applying both methods SS and SP.
The actual probability of one or more Type I errors was estimated with 
 the proportion times there were one or more rejections among
5000 replications. 
Table 2 reports the estimated probability   of one or more Type I errors when $n=20$, 
$p=4$ and 5, and
$\alpha=0.05$, 0.025 and 0.01.
The values in Table 2 for method M were taken from Wilcox (2003).

\begin{table}
\caption{Estimates of the actual FWE, $n=20$, using methods M, SP and SS}
\centering
\begin{tabular}{rcc rccc} \hline
$p$ & $g$  & $h$ & Method  & $\alpha=0.05$ & $\alpha=0.025$ & $\alpha=0.01$\\ \hline
4  & 0.0 & 0.0 & M &  0.033 & ---- & ----\\    
4  & 0.0 & 0.0 &SP & 0.058 &  0.023 & 0.011\\
4  & 0.0 & 0.0 &SS & 0.048 &  0.024 &  0.007\\
4  & 0.0 & 0.5 &M &  0.050 & ---- & ----\\
4  & 0.0 & 0.5 &SP & 0.060 & 0.031 & 0.013\\
4  & 0.0 & 0.5 &SS & 0.055 & 0.028 & 0.011\\
4  & 0.5 & 0.0 &M &  0.037 & ---- & ----\\
4  & 0.5 & 0.0 &SP & 0.064 & 0.031 & 0.013\\
4  & 0.5 & 0.0 &SS &  0.056 & 0.026 & 0.011\\
4  & 0.5 & 0.5 &M &  0.055 & ---- & ----\\
4  & 0.5 & 0.5 &SP &0.050 & 0.026 & 0.010\\
4  & 0.5 & 0.0 &SS & 0.050 & 0.023  &0.009\\

5  & 0.0 & 0.0 &M & 0.015  & ---- & ----\\
5 & 0.0 & 0.0 &SP & 0.065 & 0.030 &  0.014\\
5  & 0.0 & 0.0 &SS & 0.047 & 0.028 & 0.008\\
5  & 0.0 & 0.5 &M &  0.050 & ---- & ----\\
5  & 0.0 & 0.5 &SP & 0.049  & 0.024 & 0.008\\
5  & 0.0 & 0.5 &SS & 0.051 & 0.024 & 0.009\\
5  & 0.5 & 0.0 &M &  0.019 & ---- & ----\\
5  & 0.5 & 0.0 &SP & 0.069 &  0.034 &  0.013\\
5  & 0.5 & 0.0 &SS & 0.056 & 0.028 &  0.010 \\
5  & 0.5 & 0.5 &M &  0.054 & ---- & ----\\
5  & 0.5 & 0.5 &SP & 0.042 &  0.020 & 0.008\\
5  & 0.5 & 0.0 &SS & 0.055 & 0.028 & 0.012\\  \hline 
\multicolumn{7}{l}{M=Wilcox's  method}\\
\multicolumn{7}{l}{SP=Pearson's correlation}\\
\multicolumn{7}{l}{SS=Spearman's correlation}\\
\end{tabular}
\end{table}

Although the importance of a Type I error probability can depend on the
situation,  Bradley (1978) suggested that as a general guide, when
testing at the $\alpha$ level, the actual level should be between
$0.5 \alpha$ and $1.5 \alpha$.
As can be seen, method SS  always has estimates closer to the nominal level than method M when testing at the
0.05 level.  Moreover,  both SS and SP satisfy Bradley's criterion with 
  method SP  a bit less satisfactory than
method SS in terms of controlling FWE.

Note that when using method SP, the largest estimate in Table 2, when testing at the 0.05 level, 
occurs when $g=0.5$ and $h=0$. When $p=4$ the estimate is 0.064, and it is 0.069 when $p=5$.
Increasing the sample size to $n=40$, the estimates were 0.054 and 0.062, respectively.

Table 3 reports results using method ECP that include situations where
there is heteroscedasticty. The focus is  on $p=2$ and $n=20$, so
the results are relevant when testing (3) with method L. The number of replications 
was reduced to  2000 due to the high execution time. As a partial check on the impact of heteroscedasticty, data
were generated where $X_2=\lambda(X_1) \epsilon$, and $\epsilon$ has the same g-and-h distribution 
used to generate $X_1$. 
 Three choices for  $\lambda(X_1)$ were used:  
 $\lambda(X_1) \equiv 1$ (homoscedasticity), $\lambda(X_1)=|X_1|+1$,  or 
$\lambda(X_1)=1/(|X_1|+1)$. These three variance patterns are labeled VP 1, VP 2 and 
VP 3 henceforth.  Estimates that do not satisfy Bradley's criterion are in bold.

\begin{table}
\caption{Estimated Type I error probabilities using method ECP,  $p=2$, $n=20$.
}
\centering
\begin{tabular}{rrrr rrr} \hline
g & h &  VP  & Method  & $\alpha=0.05$ & $\alpha=0.025$ & $\alpha=0.01$  \\ \hline
  0.0 & 0.0 &  1 & HP  & 0.063 &  {\bf 0.038} & {\bf 0.017}\\
  0.0 & 0.0 &  2 & HP & 0.061 & 0.032 & {\bf 0.017}\\
   0.0 & 0.0 &  3 & HP &  0.068 & 0.030 & 0.012\\ 
  
  0.0 & 0.0 & 1 &  HS &  0.051 & 0.023 & 0.014 \\ 
   0.0 & 0.0 & 2 &  HS & 0.064  &  0.035 &   {\bf 0.022} \\    
    0.0 & 0.0 & 3 &  HS & 0.055 &  0.032 & 0.016\\   
    
   0.0 & 0.5 &  1 & HP  & 0.051 & 0.025 & 0.014\\
   0.0 & 0.5 &  2 & HP & 0.049 & 0.024 & 0.014\\  
   0.0 & 0.5 &  3 & HP &  0.053 &  0.022  & 0.011\\
     
   0.0 & 0.5 & 1 & HS & 0.050 &  0.035 & {\bf 0.019}\\
  0.0 & 0.5 &  2 &HS &  0.056 & 0.031 & {\bf 0.020} \\
  0.0 & 0.5 & 3 &HS &  0.053 &  0.030 & {\bf 0.021} \\ 
   
0.5 & 0.0 & 1& HP&  0.056 &  0.025 &  0.014 \\
0.5 & 0.0 &  2 &  HP& 0.061 &  0.029  & 0.016 \\
0.5 & 0.0 &  3 &  HP&   0.050 &  0.025 & 0.012\\

0.5 & 0.0 & 1 & HS&   0.051 & 0.032 & {\bf 0.022}\\
0.5 & 0.0 & 2&  HS& 0.061  & 0.035 & {\bf 0.024}\\
0.5 & 0.0 & 3&  HS& 0.047 & 0.027 & 0.016\\

0.5 & 0.5 & 1&  HP  &  0.050 & 0.023 & 0.014\\
0.5 & 0.5 & 2&  HP  & 0.048 & 0.023 & 0.011 \\
0.5 & 0.5 & 3&  HP  &  0.046 & 0.021 & 0.012\\

0.5 & 0.5 & 1 &  HS  & 0.047 & 0.029 & {\bf 0.016} \\
0.5 & 0.5 & 2 &  HS & 0.051 & 0.028 & {\bf 0.017}\\
0.5 & 0.5 & 3  &  HS & 0.050 & 0.028 & {\bf 0.017}\\
  \hline 
\end{tabular}
\end{table}

Note that in Table 3, given $g$ and $h$, the estimates under homoscedasticity differ very little from the estimates when
there is heteroscedasticity.  All indications are that method ECP performs reasonably well
 when testing at the 0.05 or 0.025 level. However,
 with $n=20$, control over the Type I error probability might be viewed as being unsatisfactory in some situations when testing at the 0.01 level.
But another issue is how well it performs, in terms of controlling FWE, when $p>2$.
Table 4 reports results when $p=4$ and $n=20$.
 All of the estimates satisfy Bradley's criterion.

\begin{table}
\caption{Estimates of FWE using method ECP,  $p=4$, $n=20$}
\centering
\begin{tabular}{rrr rrr} \hline
g & h &  Method  & $\alpha=0.05$ & $\alpha=0.025$ & $\alpha=0.01$  \\ \hline
  0.0 & 0.0 &  HP &  0.051 & 0.025 & 0.011\\
  0.0 & 0.0 &  HS & 0.057 & 0.030 & 0.019 \\   
  0.0 & 0.5 & HP & 0.043 & 0.024 & 0.008 \\
   0.0 & 0.5 & HS & 0.057 & 0.030 & 0.018\\
0.5 & 0.0 &  HP& 0.043  & 0.021 & 0.011\\
0.5 & 0.0 &  HS&  0.058 & 0.034 & 0.017  \\
0.5 & 0.5 &  HP  & 0.050  & 0.019 & 0.008 \\
0.5 & 0.5 &  HS  & 0.058 & 0.034 & 0.017\\  \hline 
\end{tabular}
\end{table}

Of course, controlling FWE comes at the price of reducing power. Consider the situation where $p=5$ , $n=50$ and vectors of 
observations are sampled from a multivariate normal distribution having a common  correlation of 0.3. If no adjustment aimed at
controlling FWE is made, and $H_0$: $\tau_{12}=0$ is tested at the 0.05 level using HP, power was estimated to be 0.48.
But if FWE is controlled, power is only 0.20.

Table 5 reports results for method L3
for $g=h=0$ (normality), $p=3(1)9$, and $n=30$ and 50. As can be
seen, Bradley's criterion is met for $p=3(1)8$ except 
for $p=7$ and $\alpha=0.05$, where the estimate is 
0.078.  For $p=9$, the method begins to breakdown. For $p=10$, not shown in Table 5, L3 
performs well
for $\alpha=0.05$, but the estimated level is substantially
 less than the nominal level for $\alpha=0.025$ and 0.01; the
estimates are 0.014 and 0.000 respectively.

\begin{table}
\caption{Estimates of FWE using method L3, $g=h=0$}
\centering
\begin{tabular}{rrr rrr} \hline
$p$ & $n$ & $\alpha=0.05$ & $\alpha=0.025$ & $\alpha=0.01$  \\ \hline
3 & 30  & 0.057 &  0.031 &   0.010\\
3&  50 & 0.065 &  0.040 & 0.015\\

4 & 30 & 0.057   &  0.028 &   0.018\\
4 & 50 &0.056 &  0.024 &   0.005\\

5 &  30 &  0.060 &  0.035 &  0.012\\
5 & 50 & 0.064 & 0.025 & 0.009\\

6 & 30 & 0.066 &    0.033 &     0.013\\
6 & 50 & 0.062  &    0.020 &   0.010\\

7 & 30  &0.063 &     0.038 &     0.015\\
7 & 50  &0.078  &  0.022 &  0.013\\

8 & 30 & 0.065 & 0.031&  0.018\\
8 & 50 & 0.060 &0.020 & 0.011\\

9 & 30  & 0.077 &     0.041 &     0.026\\
9 & 50  & 0.061  &  0.024 & 0.023\\ \hline

\end{tabular}
\end{table}

As for method H1, the results are similar to those in Table 5 for $p=3$ and 4.
 For $p=5$, it performs well for $\alpha=0.05$
but the estimated level is unsatisfactory, based on Bradley's criterion, for $\alpha=0.025$ and 0.01.

\section{Two Illustrations}

This section provides two illustrations. The first stems from 
Wilcox (2003) who used data   dealing with reading abilities in children  to  illustrate method M. 
The variables were a measure of digit naming speed, a measure of 
letter naming speed, and a standardized test used to measure
the ability to identify words.   
The sample size is $n=73$ after eliminating any row with missing values.  
The data are available in the file read\_dat.txt at
https://dornsife.usc.edu/labs/rwilcox/datasets/.

The estimates of Pearson's correlation are 0.106, $-0.034$ and $-0.590$.
Using the usual Student's T test with Pearson's
correlation, with no attempt
to control FWE or remove outliers, the p-values associated with 
$\rho_{12}$,  $\rho_{13}$, $\rho_{23}$, are
0.37, 0.76 and $<0.001$, respectively, indicating
a significant, negative association between a measure of 
letter naming speed and a standardized test used to measure
the ability to identify words. 
Using method M with Pearson's correlation, 
the test statistics were $|T_{12}|=8.13$,
$|T_{13}|=3.18$, and $|T_{23}|=5.62$,
with an estimated critical value of $q=2.9$.
The skipped correlations are estimated to be
0.69, $-0.35$ and $-0.55$, respectively, all three of which are significant at the 0.01 level. 
 As is evident, the first two correlations
 differ substantially from the situation where Pearson's correlation is used with the outliers included.
 
 As for  Spearman's rho, the estimates when outliers are included  are 0.453, $-0.269$ and $-0.602$,
all three of which are significant at the 0.028 level based on a percentile bootstrap method that
allows heteroscedasticity (e.g., Wilcox, 2017a). Using method M instead, 
 the corresponding estimates  are
 0.72, $-0.41$ and $-0.54$. Note that the first estimate differs substantially from the estimate of 
 Spearman's rho when the outliers are retained, which
 illustrates that outliers can have a substantial impact on Spearman's rho. 
 And there is a moderately large difference between the
 second two estimates.  Again, using M, 
all three hypotheses given by (2) are significant when testing at the 0.01 level. Method
 SS (Spearman's rho) also rejects all three when testing at the 0.05 level.
 Even when FWE is set equal to 0.0011, one association is still indicated.






 Method M indicates that there are associations, but as previously explained, it can be
 unsatisfactory when the goal is to make inferences about the skipped correlation. Using instead method ECP, 
 based on Pearson's correlation, the three p-values
 are now  $<$0.001,  0.028 and $<$0.001, respectively. 
 The Hochberg adjusted p-values stemming from H1 are 0.002,  0.020 and  0.002.
 But for method ECP, the critical p-value was estimated to be 0.026. 
 So the second hypothesis 
 is not quite significant at the 0.05 level, illustrating that in some situations, Hochberg's method might offer
 a power advantage. 
 For the situation at hand, the critical p-value used by H1 is 0.0765. 
 That is, it uses Hochberg's method by testing at the 0.0765 level, which results
 in FWE approximately equal to 0.05. Note that 
  the final step using
 Hochberg's method would  be to test at the $0.0765/3=0.0255$ level. That is, in this particular situation,
 Hochberg's method will have as much or more power than method ECP.
 Put another way, method ECP corresponds to using 
 the  Bonferroni method at the $3 \times 0.026=0.078$ level, which
 is approximately equal to the level used by H1, which is  0.0765.  And it is well-known that Hochberg's method has as much or  more  power
 than the  Bonferroni method. 

The second illustration stems from a study by Bieniek et al. (2016) 
who reported EEG data from a cross-sectional sample of 120 participants, aged 18-81, who
categorized images of faces and textures. 
Based on these data, an integrative measure of visual processing
speed was computed, using the approach described in Rousselet et al. (2010). A 
basic  issue 
was understanding the association between the dependent variable (processing speed) and age.
Two other independent variables are included here: 
 years of education and visual acuity.
Pearson's correlation between processing speed and age, education and visual acuity  was
estimated to be 0.573,
$-0.247$ and $-0.309$, respectively.
 The corresponding p-values were
$<0.001$, 0.006 and 0.001.
Using method L3 to control FWE, 
again all three independent variables have a significant association with the
dependent variable. The Hochberg adjusted p-values ranged between 0.0021 and 0.0028.
The skipped correlations were 0.621, $-0.280$ and $-0.407$. So all three skipped correlations
indicate a stronger association versus Pearson's correlation.

\section{Concluding Remarks}

As was demonstrated, methods SS and SP address the three concerns associated with method M. However,
not all 
practical limitations have been eliminated. First, computational issues arise when using methods SS and SP 
with $n=10$. Using the projection-type outlier detection method based on a bootstrap sample can result in 
MAD or the interquartile range being zero.  That is, a projection-type outlier detection method cannot
be applied due to division by zero.  Even with a large sample size, 
if there are many tied values, this issue might
occur. 

As for testing (3), all indications  are that method L3 is reasonably satisfactory,
   in terms of controlling Type I error probabilities,  when $p \le 8$. For
$p=9$ and 10, all indications that it continues to perform reasonably well when $\alpha=0.05$, but it  can be 
unsatisfactory when testing at the 0.025 or 0.01 levels. 
 As for testing (1), simulations indicate that
method H1 is adequate when $p \le 4$. Note that for $p=5$, ten hypotheses are being tested and that
results related to method L3 suggest that it will not perform well when testing at the 0.025 and 0.01
levels. Simulation results confirm this. In contrast, methods SS and SP continue to perform 
reasonably well when $p=5$.

There are several other outlier detection methods that deserve serious consideration (e.g., Wilcox, 2017a).
A few simulations were run where the  projection-type outlier detection method was replaced by the 
method derived by Rousseeuw and van Zomeren (1990), as well as a method based in part on the
minimum covariance determinant estimator (e.g., Wilcox, 2017a, section 6.3.2).
Results were very similar to those presented here, but a more definitive study is needed.

Finally, the R function mscorpb  applies SP by default, which is aimed at testing (2). 
Method SS can be applied by setting the argument corfun=spear. The
function contains an option for using alternative outlier detection techniques.  As for testing (1), 
the function mscorci applies
method H by default in conjunction with Pearson's correlation. Setting the argument hoch=FALSE, method ECP is used.  Setting the argument corfun equal to spear, Spearman's rho is used.
To reduce execution time, method H1 can be used via the R function mscorciH. 
As for testing (3), the R function scorregci can be used. If the goal is merely to estimate the skipped correlations without
testing  hypotheses, use the R function scorreg. 
 A Matlab implementation of the skipped correlation is also available 
 (Pernet, et al., 2013.) The R function scorregciH applies method L3. 
All of these functions are being added to the
R package WRS. For a more detailed description of how the data in section 6 were analyzed
using these functions, 
see Wilcox et al. (2018).
 
\begin{center}
References
\end{center}

Bieniek, M. M.,  Bennett, P. J.,
Sekuler, A. B.  \&  Rousselet, G. A. (2016). A robust and representative lower bound on
object processing speed in humans. 
{\em European Journal of Neuroscience, 44}, 
1804--1814.

Bradley, J. V. (1978). Robustness? {\em British Journal of Mathematical and Statistical Psychology, 31}, 144--152.

Carling,  K. (2000). Resistant outlier rules and the non-Gaussian case.
 {\em Computational Statistics \& Data Analysis, 33}, 249--258.
 

Croux, C. \& Dehon, C. (2010). Influence functions of the Spearman and Kendall
Correlation measures. {\em Statistical Methods and Applications, 19}, 497--515.

Devlin, S. J., Gnanadesikan, R. \& Kettenring, J. R. (1981). Robust estimation
 of dispersion matrices and principal components. {\em Journal of the
  American Statistical Association, 76}, 354--362.
 
Donoho, D. L. (1982). Breakdown properties of multivariate location estimators. PhD
qualifying paper, Harvard University

Harrell, F. E. \& Davis, C. E. (1982). A new distribution-free quantile 
 estimator. {\em Biometrika, 69}, 635--640.
 
Hoaglin, D. C. (1985) Summarizing shape numerically: The g-and-h 
distributions. In D. Hoaglin, F. Mosteller \& J. Tukey (Eds.)
{\em Exploring Data Tables, Trends and Shapes}. New York: Wiley.

Hochberg, Y. (1988). A sharper Bonferroni procedure for multiple tests of 
 significance. {\em Biometrika, 75}, 800--802.


Liu, R. G. \& Singh, K. (1997). Notions of limiting P values based on data
 depth and bootstrap. {\em Journal of the American Statistical Association, 92},
 266--277.
 
 Pernet, C. R., Wilcox, R. \& Rousselet, G A. (2013). Robust correlation analyses:
  a Matlab toolbox for psychology research. {\em Frontiers in Quantitative Psychology
  and Measurement}.  DOI=10.3389/fpsyg.2012.00606
 
Rousseeuw, P. J. \& van Zomeren, B. C. (1990). Unmasking multivariate outliers
 and leverage points. {\em Journal of the American Statistical Association},
 {\em 85}, 633--639.

Rousselet,
G.  A., Gaspar, C. M.,  Pernet, C. R., Husk, J. S.,  Bennett, P. J. \& Sekuler,
A. B. (2010). Healthy aging delays scalp
EEG sensitivity to noise in a face discrimination task.
{\em Frontiers in psychology, 1}.


Stahel, W. A. (1981). Breakdown of covariance estimators. Research report 31, Fachgruppe f\"{u}r  Statistik, E.T.H Zurich.


Wilcox, R. R. (2003). Inferences based on multiple skipped correlations. {\em Computational
Statistics \& Data Analysis, 44}, 223--236.

Wilcox, R. R  (2015). Inferences about the skipped correlation coefficient:
Dealing with heteroscedasticity and non-normality. {\em Journal of Modern}
 {\em and Applied Statistical Methods, 14}, 2--8.  

Wilcox, R. R. (2017a). {\em Introduction to Robust Estimation and Hypothesis Testing}, 4th Ed.  San Diego, CA: Academic Press.

Wilcox, R. R. (2017b). {\em Understanding and Applying Basic Statistical Methods Using R}. New York: Wiley.

Wilcox, R., Pernet, C. \& Rousselet,  G. (2018). Improved methods for making inferences about multiple skipped correlations.
 figshare. https://doi.org/10.6084/m9.figshare.5768301.v1


\end{document}